\documentclass{article}
\usepackage{spconf,amsmath,graphicx}
\usepackage{bbold}
\usepackage[normalem]{ulem}
\usepackage{hyperref}
\usepackage[dvipsnames]{xcolor}
\usepackage{amsmath}
\usepackage{subfigure}
\usepackage{url}
\usepackage{appendix}
\usepackage{enumitem}
\makeatletter
\def\endthebibliography{%
	\def\@noitemerr{\@latex@warning{Empty `thebibliography' environment}}%
	\endlist
}
\makeatother

\title{Real-World Adversarial Examples involving Makeup Application}
\name{Chang-Sheng Lin$^1$, Chia-Yi Hsu$^2$, Pin-Yu Chen$^3$, Chia-Mu Yu$^2$
\address{Nation Chung Hsing University$^1$\\
National Yang Ming Chiao Tung University$^2$\\
IBM Thomas J. Watson Research Center$^3$}
}
\begin{document}
\ninept
\maketitle
\begin{abstract}
Deep neural networks  have developed rapidly and have achieved outstanding performance in several tasks, such as image classification and natural language processing. However, recent studies have indicated that both digital and physical adversarial examples can fool  neural networks. Face-recognition systems are used in various applications that involve security threats from physical adversarial examples. Herein, we propose a physical adversarial attack with the use of full-face makeup. The presence of makeup on the human face is a reasonable possibility, which possibly increases the imperceptibility of attacks. In our attack framework, we combine the cycle-adversarial generative network (cycle-GAN) and a victimized classifier. The Cycle-GAN is used to generate adversarial makeup, and the architecture of the victimized classifier is VGG 16. Our experimental results show that our attack can effectively overcome manual errors in makeup application, such as color and position-related errors. We also demonstrate that the approaches used to train the models can influence physical attacks; the adversarial perturbations crafted from the pre-trained model are affected by the corresponding training data.
\end{abstract}

\section{Introduction}
Deep neural networks are well-known for their impressive performance in machine learning and artificial intelligence applications, such as object detection, automatic speech recognition, and visual art processing. However, recent research has demonstrated that well-trained deep neural networks are vulnerable to indistinguishable perturbations called adversarial examples, which can be applied in both digital and physical attacks. Extensive efforts have been devoted to addressing digital adversarial attacks. Madry et al.\cite{madry2017towards} proposed an iterative gradient-based attack that can effectively search for adversarial examples within the allowed norm ball. Carlini and Wanger \cite{carlini2017towards} formalized adversarial attacks as an optimization problem and found imperceptible perturbations. Moreover,
an ample set of digital attacks (\cite{deng2019arcface, dong2019efficient,deb2019advfaces,zhu2019generating,zhong2020towards,yang2021attacks}) can craft unnoticeable and strong perturbations over the entire image against face recognition (FR) systems. In practice, however, digital attacks cannot be directly applied in the physical world. For instance, in the setting of digital attacks, the malicious attacker attacking FR without any restriction for the positions of adversarial perturbations against the actual situation.  In a reasonable scenario, a malicious attacker attempting to mislead the FR system can only add perturbations to the face instead of the background. Thus, a physical attack, which has more limitations than a digital attack, is more complicated. In addition to the positions of perturbations, adversarial perturbations are affected by several environmental factors, such as brightness, viewing angle, and the camera resolution in physical attacks. There have also been several efforts to address physical attacks. Certain physical attacks \cite{sharif2016accessorize, xu2020adversarial,komkov2021advhat} have overcome specific limitations associated with printing adversarial noise on wearable objects, such as eyeglasses, T-shirts, and hats. Moreover, some studies have focused on attacking FR systems using adversarial patches \cite{pautov2019adversarial} and adversarial light \cite{nguyen2020adversarial}. All these studies considered environmental factors and the reducibility of adversarial perturbations.
\begin{figure}[t]
    \centering
    \includegraphics[width=0.49\textwidth]{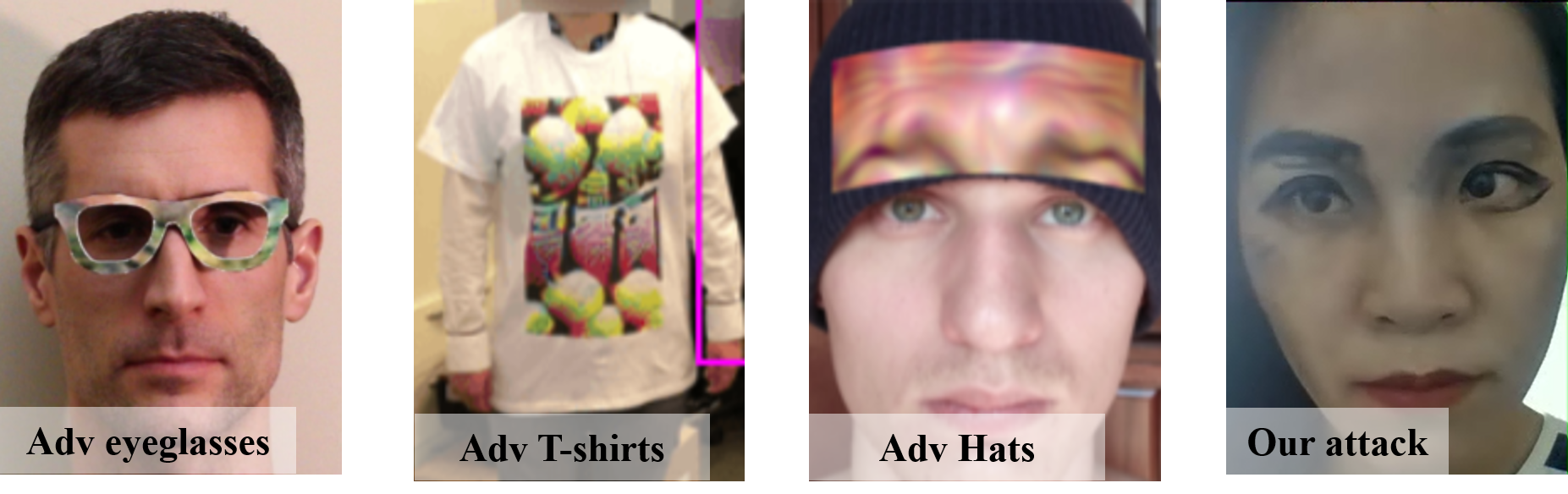}
    \caption{Illustration of physical adversarial examples generated by Adv eyeglasses, Adv T-shirts, Adv Hat, and our attack.}
    \label{fig:comparison of physical attack}
\end{figure}

In this study, inspired by \cite{zhu2019generating}, we designed an attack that uses full-face makeup as adversarial noise. Instead of printing, we aimed to manually perturb the face and ensure that it would mislead the FR system successfully. Compared with prior work on physical attacks, the most notable difference, and also the most challenging aspect, of our attack is the method of reproducing the noise from digital results. As shown in Fig.\ref{fig:comparison of physical attack}, the adversarial examples crafted under the physical adversarial examples crafted in prior studies are visually distinctive to the human eye, whereas our adversarial example has a more natural appearance. Our contributions are summarized as follows: (1) We propose a novel method for synthesizing adversarial makeup. (2) When implemented in the real world, our attack can compensate for manual errors in makeup application and is thus an example of an effective physical adversarial example.

\section{Related Work}

\subsection{Adversarial Attacks}
Adversarial attacks can be conducted using digital and physical methods. Digital attacks involve fewer restrictions than physical attacks. In the physical scenario, many factors affect the presentation of adversarial perturbations, such as the light and angle of the camera lens. Both digital and physical attacks can be defined as targeted and untargeted attacks. The definition of a targeted attack is stricter, that is, the prediction result of the adversarial example must be a specific class. However, the output of the model is only different from the ground truth label in an untargeted attack. We present the details of digital and physical attacks in the following sections.

\subsubsection{Digital Attacks}
Several studies on attack methods have recently demonstrated that deep neural networks (DNNs) can be easily fooled by adversarial examples. In general, the loss function of a digital adversarial attack comprises the restrictions on perturbations and attack loss. For instance, Szegedy et al. \cite{szegedy2013intriguing} proposed that given an input $x$, one can find a solution $r$ that allows the classified result of $x+r$ to be close to the target class and $r$ to be small. This can be formalized as an optimization problem:
\begin{equation}\label{eq:digital loss}
    \mathop{{\rm minimize}}\limits_{r} \quad c \vert r \vert + \mathcal{L}(f(x+r), \ell_t),\; \text{subject to}\; x+r \in [0,1]^m
\end{equation}
where $\mathcal{L}$ is a function to compute the distance between two probability distributions, such as the cross-entropy, $f(\cdot)$ is the victimized model, $\ell_t$ is the target label, and $m$ denotes the data dimension. The hyper-parameter $c$ governs the importance of the norm of perturbations $r$. In addition to optimization-based attacks, Goodfellow et al. \cite{goodfellow2014explaining}, Madry et al. \cite{madry2017towards},
and Dong et al. \cite{dong2018boosting} proposed gradient-based methods to attack DNNs.

Based on the purpose of out attack, we introduce several digital attacks on FR systems in this section. Zhu et al. \cite{zhu2019generating} first attempted to use eye makeup to perturb the target input and then attack the FR system. Yang et al. \cite{yang2021attacks} used a generative neural network to generate adversarial face images that attack an FR system. Adversarial examples generated using these approaches (\cite{deng2019arcface, dong2019efficient,deb2019advfaces,zhong2020towards}) either appear factitious or cannot be directly applied in the physical world.
\subsubsection{Physical Attacks}
A physical attack requires more factors to be considered, and it uses an objective function similar to that in digital attacks. Considering Eq.\ref{eq:digital loss}, however, the constraint on $r$ is not sufficient, which results in the failure of the physical attack . Sharif et al. \cite{sharif2016accessorize} suggested that there are three aspects that should be considered for perturbations of $r$: (1) how perturbations can be added in the real world; (2) environmental factors: light, positions of adversarial noise, and angle of the camera lens; (3) increasing the smoothness of the adversarial noise. Accordingly, they proposed a patch-based attack to add perturbations within a specific region, e.g., the area covered by eyeglasses, to attack FR systems. Similar attacks on wearable objects were also synthesized by \cite{xu2020adversarial, komkov2021advhat}. Yin et al. \cite{yin2021adv} proposed Adv-Makeup, which transfers eye makeup to perform attacks with a black-box setting .
\subsection{Cycle-GAN}
\begin{figure}[h]
    \centering
    \includegraphics[width=0.45\textwidth]{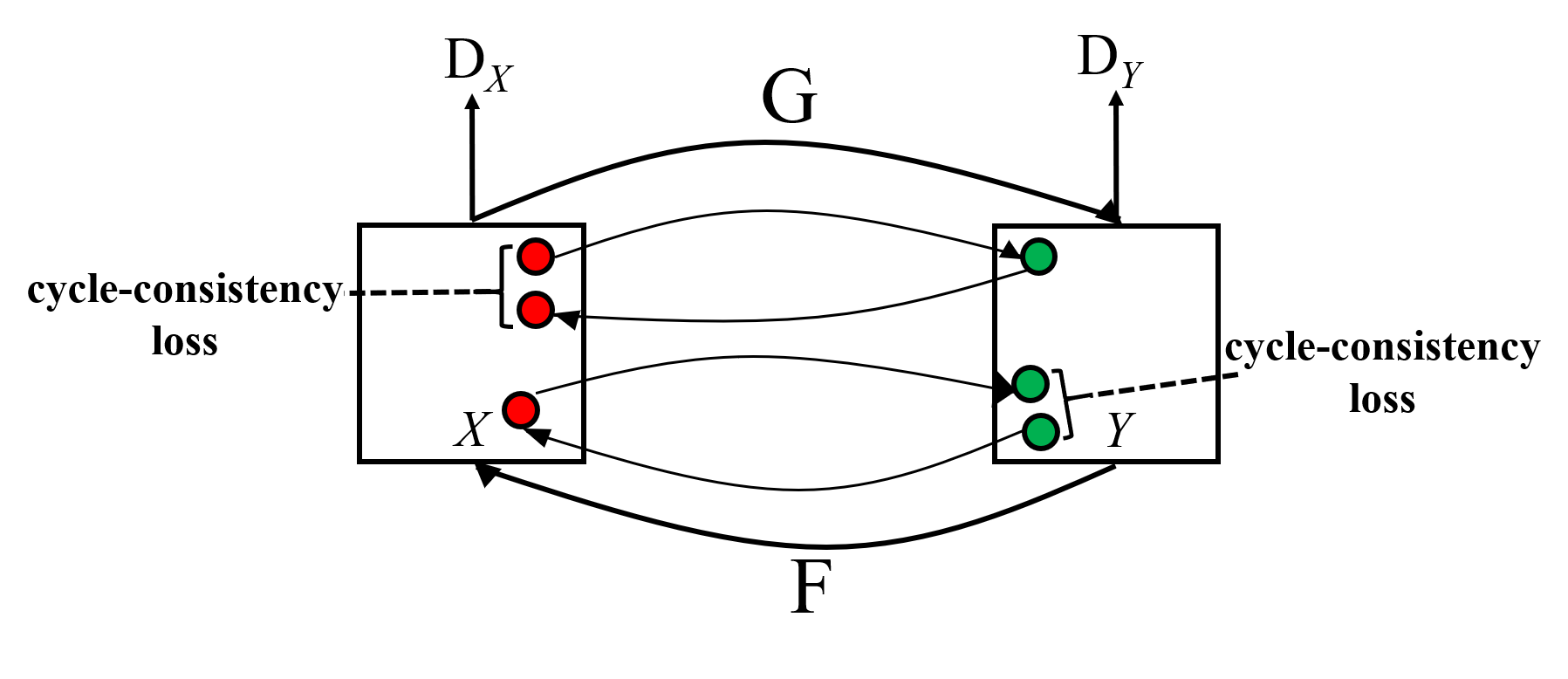}
    \vspace{-2mm}
    \caption{Framework of Cycle-GAN. It is composed of two mapping functions, G: $X \rightarrow Y$ and F: $Y \rightarrow X$, and two associated discriminators, $\text{D}_X$ and $\text{D}_Y$. $\text{D}_X$ is used to increase the similarity between the synthetic image from G and domain $Y$ and vice versa for F and $\text{D}_Y$. The cycle-consistency loss can force G and F to be consistent with each other.}
    \label{fig:cycle_gan}
\end{figure}
Cycle-GAN \cite{zhu2017unpaired} is a technique that involves unsupervised training of an image-to-image translation model with unpaired examples. Its applications include style transfer, object transfiguration, season translation, and generation of photographs from paintings. As shown in Fig.\ref{fig:cycle_gan}, Cycle-GAN comprises mapping functions and discriminators and aims to learn the mapping functions between two domains $X$ and $Y$, given training sets $\{x_i\}_{i=1}^N \in X$ and $\{y_k\}_{k=1}^M \in Y$. Its objective function contains forward--backward adversarial losses and a cycle-consistency loss, which allow images to be translated into other styles. Considering the applications of Cycle-GAN, it can be used effectively for our attack, which involves transferring images of faces both with and without makeup.

\section{Methodology}
\begin{figure*}[t]
    \centering
    \includegraphics[width=\textwidth]{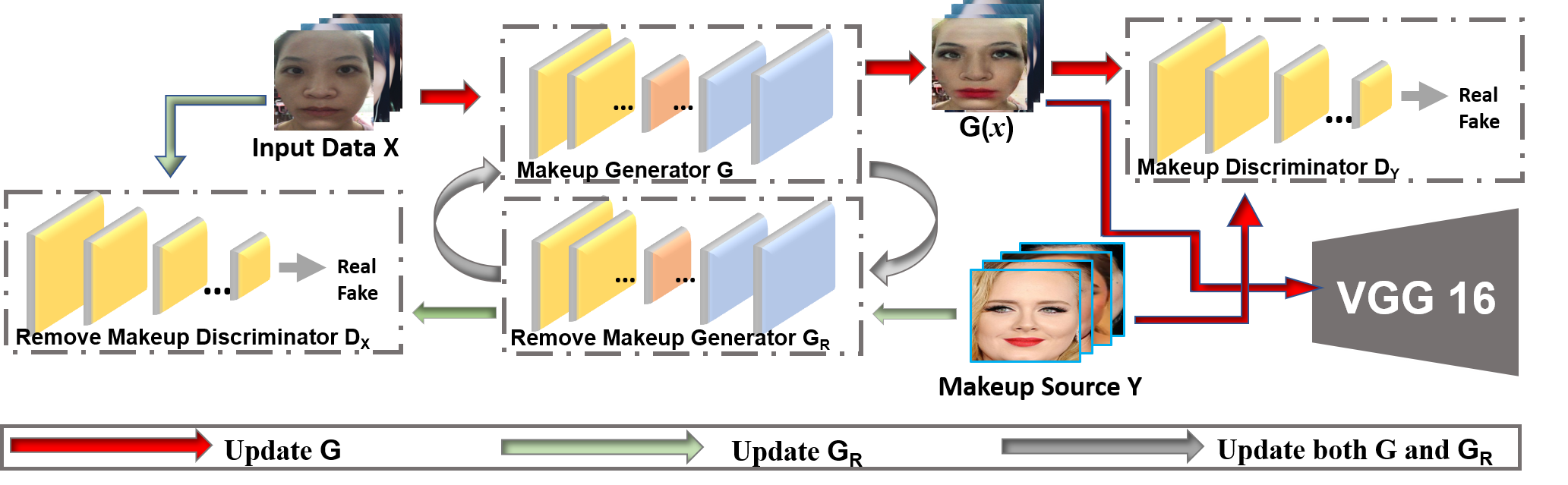}
    \caption{Overview of the framework of our attack, which consists of Cycle-GAN and the VGG 16 face-recognition classifier. Cycle-GAN contains two generators ($G$ and $G_\text{R}$) and discriminators ($D_X$ and $D_Y$). Among them, generator $G$ can generate adversarial faces with full-face makeup, successfully misleading the face recognition (FR) system, VGG 16.}
    \label{fig:makeup_framework}
\end{figure*}
\subsection{Overview}
We used the Cycle-GAN framework to generate imperceptible adversarial examples. Instead of adding irrelevant noise to the images, full-face makeup is used as adversarial perturbation to mislead well-trained FR systems. As shown in Figure \ref{fig:makeup_framework}, the framework consists of two components. One is the architecture of Cycle-GAN, which is responsible for translating the image styles between those with and without makeup. The other is the victimized FR classifier, VGG 16. With images of an individual not wearing makeup as the input data and randomly selecting faces with cosmetics applied, the makeup generator can synthesize a face with full-face makeup, misleading the VGG 16 successfully. When the makeup generator has been trained, randomly selected non-makeup images of the same individual with the input data can fool the face recognition system, VGG 16.
\subsection{Makeup Generation}
The purpose of our attack is to generate unobtrusive adversarial examples. Considering applications in the physical world, full-face makeup, which provides assorted appearances and is common in daily life, can be enforced easily. To achieve this goal, we selected Cycle-GAN, which involves automatic training of image-to-image translation models without paired examples. As shown in Figure \ref{fig:makeup_framework}, we follow the setting of Cycle-GAN \cite{zhu2017unpaired}, which comprising two generators and two discriminators. Cycle-GAN contains two GAN architectures. The makeup generator $G$ translates non-makeup images to full-face makeup images, and generator $G_\text{R}$ can transform images that contain makeup to non-makeup images. The discriminator $D_Y$ is used to stimulate the perceptual authenticity of the synthetic image featuring cosmetics, and $D_X$ is applied to improve the quality of the generative image 
reconstructed by $G_\text{R}(\cdot)$.

With the input of the non-makeup source image $x \in X$ and makeup image $y \in Y$, we first employ face detection using YoLov4 to perform face cropping for input $X$. Considering that FR classifiers are used in real life, YoLov4 should correctly classify faces with different angles to obviate the need for face alignment. The generator $G$ takes non-makeup images as input and outputs $G(\cdot)$ with generative full-face makeup; the generator $G_\text{R}$ takes $G(\cdot)$ as input and outputs $G_\text{R}(\cdot)$ without cosmetics. To improve the quality of the synthetic images, we also applied discriminators that cause the synthetic images to appear more natural. The discriminator $D_Y$ takes the real source image with cosmetics and the output $G(\cdot)$ with generative full-face makeup from the generator $G$ as input, and the discriminator $D_X$ takes the real non-makeup source image and the output $G_\text{R}(\cdot)$ without makeup generated by the generator $G_\text{R}$ as input. Cycle-GAN contains two GAN networks; thus, we define the loss of GAN as follows:
\begin{align}
    \mathcal{L}_{\text{GAN}}(G, G_\text{R}, D_X, D_Y, X , Y)& = \mathbb{E}_{y \sim p_{Y}} [\log D_Y(y)] \nonumber\\
    &+ \mathbb{E}_{x \sim p_{X}} [\log (1 - D_Y(G(x)))] \nonumber\\
    &+\mathbb{E}_{x \sim p_{X}} [\log D_X(x)] \nonumber\\
   & +\mathbb{E}_{y \sim p_{Y}} [\log (1 - D_X(G_{\text{R}}(y)))]
\end{align}
To ensure consistency between $G$ and $G_\text{R}$, $x\rightarrow G(x) \rightarrow G_\text{R}(G(x)) \approx x$, and vice versa, the loss $\mathcal{L}_\text{cycle}$ is defined as
\begin{align}
    \mathcal{L}_\text{cycle} (G, G_\text{R}, X, Y)&= \mathbb{E}_{x\sim p_{X}}[\vert\vert G_\text{R}(G(x)) - x \vert\vert_{1}] \nonumber\\
    & + \mathbb{E}_{y\sim p_{Y}}[\vert\vert G(G_\text{R}(y)) - y \vert\vert_{1}].
\end{align}
Furthermore, we introduce the loss $\mathcal{L}_\text{identity}$ to limit the differences between the input and output of the generators. $\mathcal{L}_\text{identity}$ is formalized as follows:
\begin{align}
    \mathcal{L}_\text{identity} (G, G_\text{R}, X, Y)&= \mathbb{E}_{x\sim p_{X}}[\vert\vert G_\text{R}(x) - x \vert\vert_{1}] \nonumber\\
    & + \mathbb{E}_{y\sim p_{Y}}[\vert\vert G(y) - y \vert\vert_{1}].
\end{align}
Therefore, the full objective of the Cycle-GAN is
\begin{align}
    \mathcal{L}_\text{Cycle-GAN}(G, G_\text{R}, D_X, D_Y, X, Y) &= \mathcal{L}_\text{GAN}(G, G_\text{R}, D_X, D_Y, X, Y) \nonumber \\
    &+ \lambda \mathcal{L}_\text{cycle}(G, G_\text{R}, X, Y) \nonumber\\
    &+ \alpha \mathcal{L}_\text{identity}(G, G_\text{R}, X, Y),
\end{align}
where $\lambda$ and $\alpha$ govern the corresponding importance of other objectives.

\subsection{Makeup Attack}
The most difficult aspect of using makeup as an adversarial perturbation is that people cannot apply  makeup precisely. Manual application of makeup on the face cannot exactly the match the digital result. To overcome this challenge, we use \textbf{ Gaussian blur}, denoted as $\Phi(\cdot)$, which can dim the boundaries of the makeup to simulate manual errors. Then, to produce the makeup-based adversarial perturbations, we introduce the following untargeted attack objective function:
\begin{equation}\label{eq:untargeted attack}
\mathcal{L}_\text{adv} =  \max \{ {\rm Z}(x)_{l_{x_0}} - \max( {\rm Z} (x)_{i: i \neq  l_{x_0}} ), -\kappa\}.
\end{equation}
Let $x = \Phi( G(x_0 + \delta))$ denote the Gaussian blur output of the perturbed example of $x_0$, subject to $x \in [0,1]^d$, where $d$ is the data dimension, and $[0,1]$ denotes the space of valid data examples. ${\rm Z}(x)$ is the output of x in the pre-softmax layer (known as logits), and $l_{x_0}$ is the ground-truth label of $x_0$. $\kappa \geq 0$ is a hyper-parameter that controls the model confidence of $x$. If $\kappa$ is set higher, the adversarial example will have a stronger classification confidence. The targeted attack loss can be defined as a similar loss from Eq. (\ref{eq:untargeted attack}).

In summary, we solve the optimization problem to minimize the loss function $\mathcal{L}_\text{total}$. We summarize our complete attack loss function $\mathcal{L}_\text{total}$, which combines Cycle-GAN and generates adversarial examples, as follows:
\begin{equation}
    \mathcal{L}_\text{total} = \mathcal{L}_\text{Cycle-GAN}(G, G_\text{R}, D_X, D_Y, X, Y) + \mathcal{L}_\text{adv}.
\end{equation}
\section{Experiment}
We obtained the results of our attack in a white-box setting and performed both untargeted and targeted attacks. We collected a non-makeup image dataset, which consists of images of eight colleagues from our laboratory. There were 2286 images in the training set and 254 samples in the test set. We used the makeup dataset employed by Chen et al. \cite{chen2017spoofing}, which contains 361 training samples. Our experimental results showed that the prediction probability for each class is calculated using the following equation:
\begin{equation}
    \text{P}_i = \frac{\text{the number of the frames are classified to Class}\; i}{\text{the total number of frames of the video}} \times 100\%
\end{equation}
where $\text{P}_i$ denotes that the percentage of frames is classified as Class $i$.
\begin{figure*}[t]
    \centering
    \includegraphics[width=0.95\textwidth]{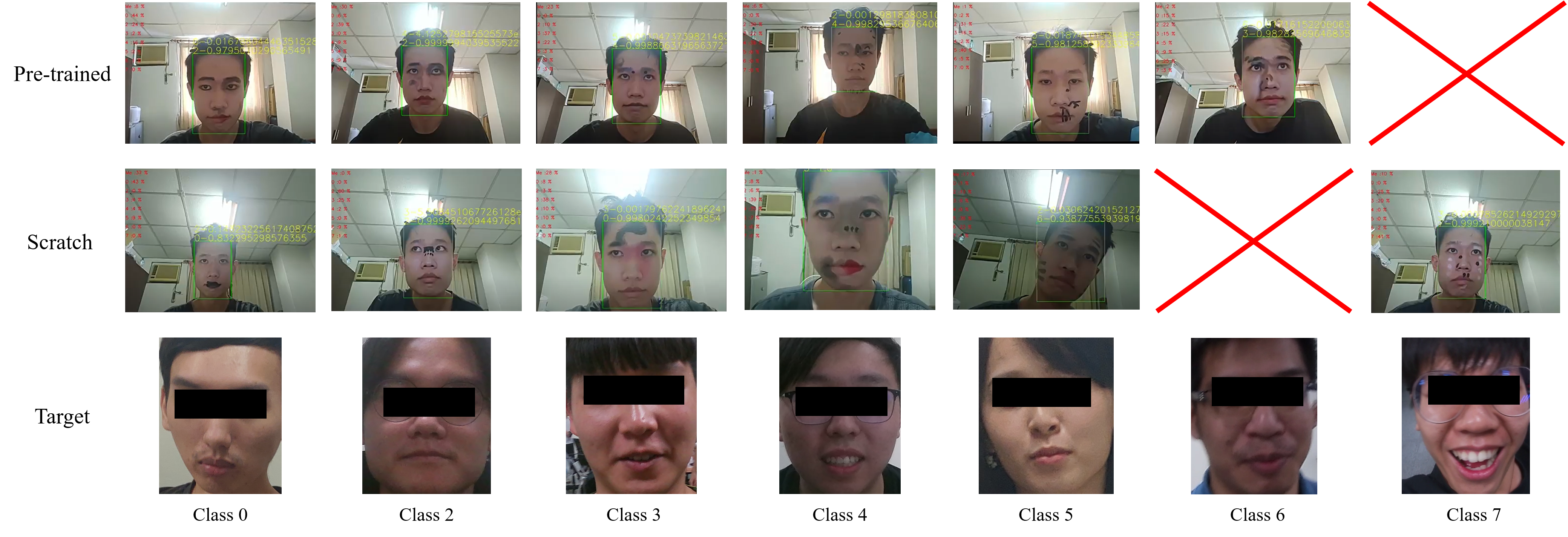}
    \caption{Visual comparison of adversarial examples generated by attacking models trained with pre-trained weights and from scratch under the setting of the targeted attack. The red crosses indicate that physical attacks failed. The targeted class is numbered 0, 2, 3, 4, $\dots$, 7 from left to right (the attacker is class 1). }
    \label{fig:targeted_attack}
\end{figure*}
\subsection{Experiment Setup}
We conducted untargeted and targeted attacks in a white-box setting, meaning that attackers could access all parameters of the model. For the coefficients of our attack objective function, we set $\alpha = 50$, $\lambda = 100$, and $\kappa =5$. We trained the classifier from the pre-trained weights and scratches. For the training with pre-trained weights\footnote{https://github.com/rcmalli/keras-vggface}, we selected Adam as the optimizer, trained the model with 367 epochs, and set the learning rate to 0.00001. For the training from scratch, we used the Adam optimizer with a learning rate of 0.00001 and 408 epochs. For both training methods, we set the batch size to 25. In our attack, we used the Adam optimizer with a learning rate of 0.0002 and set the batch size to 1. We ran our attack with more than 100 epochs and then selected the images that appeared the most natural as the adversarial examples. All the experiments  were conducted using a PC with an Intel Xeon E5-2620v4 CPU, 125 GB RAM, and an NVIDIA TITAN Xp GPU with 12 GB RAM. The camera used was an ASUS ZenFone 5Z ZS620KL (rear camera).

\subsection{Untargeted Attack}
\begin{figure}[h]
    \centering
    \includegraphics[width=0.48\textwidth]{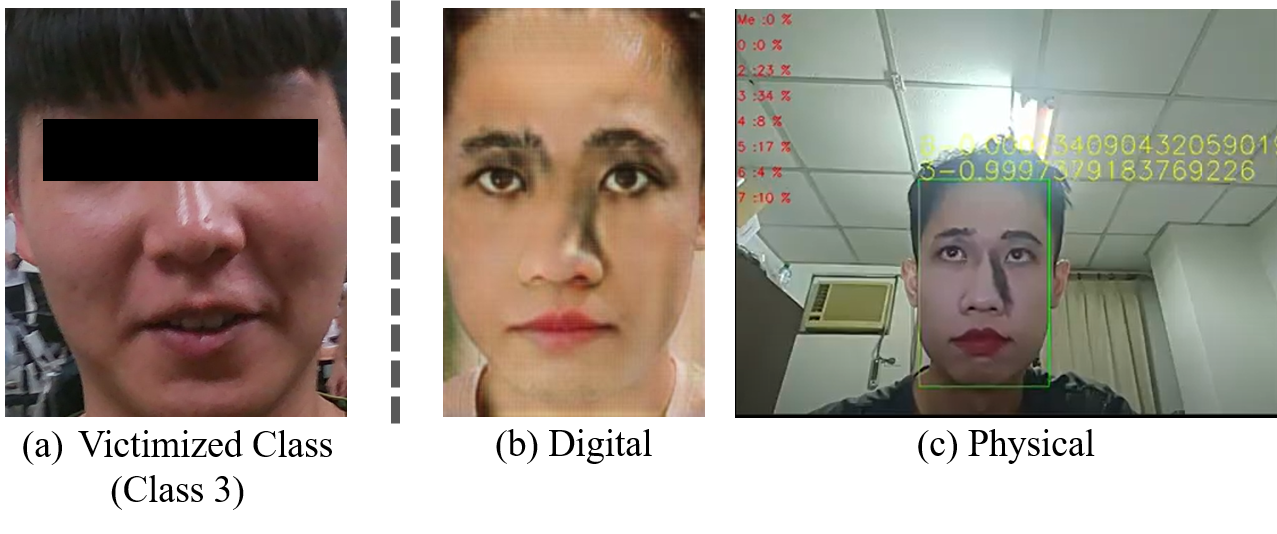}
    \caption{Visual comparison of physical and digital adversarial examples generated under the setting of the untargeted attack. (a) showed the person  who is classified when taking physical adversarial examples as the input. (b) Digital adversarial sample generated by the attack. (c) Result of an attacker wearing makeup.}
    \label{fig:untargeted}
\end{figure}

\begin{figure}[h]
    \centering
    \subfigure[Pre-trained]{
    \includegraphics[width=0.23\textwidth]{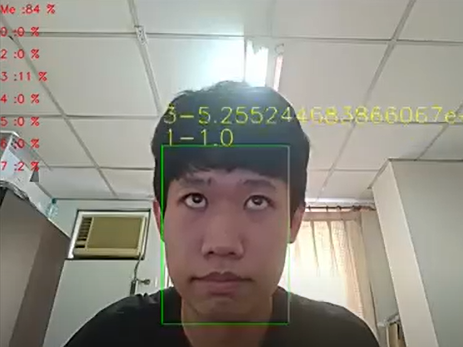}}
    \subfigure[Scratch]{
    \includegraphics[width=0.23\textwidth]{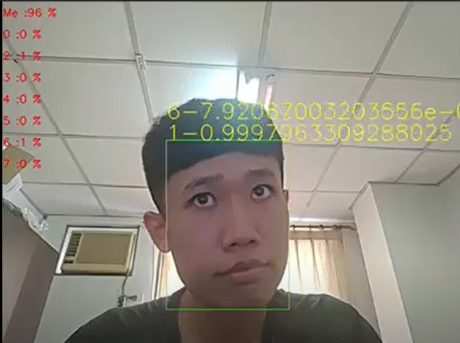}
    }
    \caption{(a) shows that the attacker can be classified to himself with the 84\% when using the model trained from the pre-trained weights. (b) showed the attacker had 96\% to be classified to himself by the model trained from the scratch.}
    \label{fig:normal-tran_vs_non-tran}
\end{figure}

Under an untargeted attack, the classifier trained with the pre-trained weights achieved an accuracy of 98.41\% on the test set. In the physical world, the accuracy of the attack could reach 84\%, as shown in Fig. \ref{fig:normal-tran_vs_non-tran} (a). As shown in Fig. \ref{fig:untargeted} (c), the accuracy of the attacker reduces to 0\% and the attacker has 34 percentage to be classified to the Class 3. The person in Class 3 (victimized class) shown in Fig.  \ref{fig:untargeted} (a). Fig. \ref{fig:untargeted} (b) and (c) show that the physical adversarial example is not identical to the digital one. However, it can still attack successfully when the adversarial noise is reduced.


\subsection{Targeted Attack}
We trained the classifiers with pre-trained weights and from scratch on the targeted attack. The model trained using the pre-trained weights attained an accuracy of 98.41\% on the test set. In addition, the accuracy of the model trained from scratch on the test set was 97.64\%. In the physical setting, the attack achieves accuracies of 84\% and 96\% with the pre-trained model and the model trained from scratch, respectively, as shown in Fig. \ref{fig:normal-tran_vs_non-tran}. The model trained from scratch is more robust; hence, the attacker can be classified correctly even when the viewing angle is varied. In Fig.\ref{fig:targeted_attack}, however, the attacker can get the higher percentage of some targeted classes as attacking the model trained from the scratch.  Moreover, if the targeted images have prominent features such as eyeglasses, they might be presented in the adversarial examples as well.

\section{Conclusion}
In this paper, we proposed a novel and powerful attack mechanism for real-world applications, which can utilize full-face makeup images to perform attacks on FR systems. Instead of adding adversarial perturbations using machines, our attack method adds them manually and overcomes errors associated with color and positions. The experimental results showed that our method is effective under the settings of both targeted and untargeted attacks. In future, we will attempt to reduce the amount of adversarial noise to make the perturbations less perceptible. We also intend to demonstrate that the method of training the models affects the physical attack.

\bibliographystyle{IEEEbib}
\bibliography{ref}

\end{document}